\documentclass{ws-procs9x6}            % default, citations in superscript
\begin{document}

\title{Recent Progress in Basis Light-front Quantization}

\author{Xingbo Zhao,$^*$ Kaiyu Fu,$^\$$ Hengfei Zhao,$^\dagger$ Jiangshan Lan,$^{\star\star}$ Chandan Mondal,$^{\$\$}$ Siqi Xu,$^\ddagger$}

\address{Institute of Modern Physics, Chinese Academy of Sciences, Lanzhou 730000, China\\
School of Nuclear Science and Technology, University of Chinese Academy of Sciences,
Beijing 100049, China\\
$^*$E-mail: xbzhao@impcas.ac.cn,$^\$$E-mail: kaiyufu@impcas.ac.cn\\
$^\dagger$E-mail: zhaohengfei@impcas.ac.cn,$^{\star\star}$E-mail: jiangshanlan@impcas.ac.cn\\
$^{\$\$}$E-mail: mondal@impcas.ac.cn,$^\ddagger$E-mail: xsq234@impcas.ac.cn\\
}

\author{James P. Vary}

\address{Department of Physics and Astronomy, Iowa State University,\\ Ames, IA 50011, USA\\
E-mail: jvary@iastate.edu\\
$($BLFQ Collaboration$)$}

\begin{abstract}
Basis Light-front Quantization (BLFQ) is a nonperturbative approach to quantum field theory. In this paper, we report our recent progress in applying BLFQ to the positronium system in QED and to the meson and the baryon system in QCD. We present preliminary results on the mass spectrum, light-front wave functions and other observables of these systems, where one dynamical gauge boson is retained for the positronium and meson systems.
\end{abstract}

\keywords{Light-front quantization; positronium; meson; baryon}

\bodymatter

\section{Introduction}

Basis Light-front Quantization (BLFQ) has been developed as a nonperturbative approach to relativistic bound states~\cite{Vary:2009gt}. It is based on the Hamiltonian formalism and the light-front quantum field theory. In BLFQ, the bound state problem is cast into an eigenvalue problem of the Hamiltonian: 
\begin{equation}
  P^{-}|\beta\rangle = P^{-}_{\beta}|\beta \rangle,
\end{equation}
where the eigenvalues $P^{-}_{\beta}$ correspond to the mass spectrum and the eigenvectors $|\beta \rangle$ encode their structural information. In this paper, we report our recent progress in applying BLFQ to the positronium system in QED and the meson and baryon system in QCD.

\section{Positronium}
The positronium (``Ps") is arguably the simplest bound state system in QED. In this work, we solve the positronium system from first principles - the QED Lagrangian~\cite{Fu:2020b}. In order to make the numerical calculation feasible, we perform basis truncation by retaining the two leading Fock sectors, that is, $|\text{Ps}\rangle= a|e^+ e^-\rangle+ b|e^+ e^- \gamma\rangle$. In addition, we truncate the basis in the transverse (longitudinal) direction with the truncation parameter $N_{\rm{max}}$ ($K$)~\cite{Vary:2009gt}. Larger $N_{\rm{max}}$ ($K$) translates to more complete bases in the transverse (longitudinal) direction. 
We obtain our light-front QED Hamiltonian from the QED Lagrangian via the Legendre transformation. In our truncated basis the light-front QED Hamiltonian, using light-front gauge, takes the following form,
\begin{equation}
\begin{aligned}
P^-_{\rm{QED}}= &\int \mathrm{d}^{2} x^{\perp} \mathrm{d} x^{-} \frac{1}{2} \bar{\Psi} \gamma^{+} \frac{m_{e0}^{2}+\left(i \partial^{\perp}\right)^{2}}{i \partial^{+}} \Psi+\frac{1}{2} A^{j}\left(i \partial^{\perp}\right)^{2} A^{j} \\ &+e j^{\mu} A_{\mu}+\frac{e^{2}}{2} j^{+} \frac{1}{\left(i \partial^{+}\right)^{2}} j^{+} ,
\end{aligned}
\label{QEDHami}
\end{equation}
where $\psi$ and $A_\mu$ are the fermion and gauge boson field operators, respectively, and $j^\mu=\bar{\Psi}\gamma^\mu \Psi$.
The first two terms are their corresponding kinetic terms and the remaining terms describe their interaction. $m_{e0}$ is the bare fermion mass. For numerical convenience, we take an artificially increased electromagnetic coupling constant $\alpha=0.075$.

\begin{figure}
\centering
\includegraphics[width=0.45\columnwidth
]{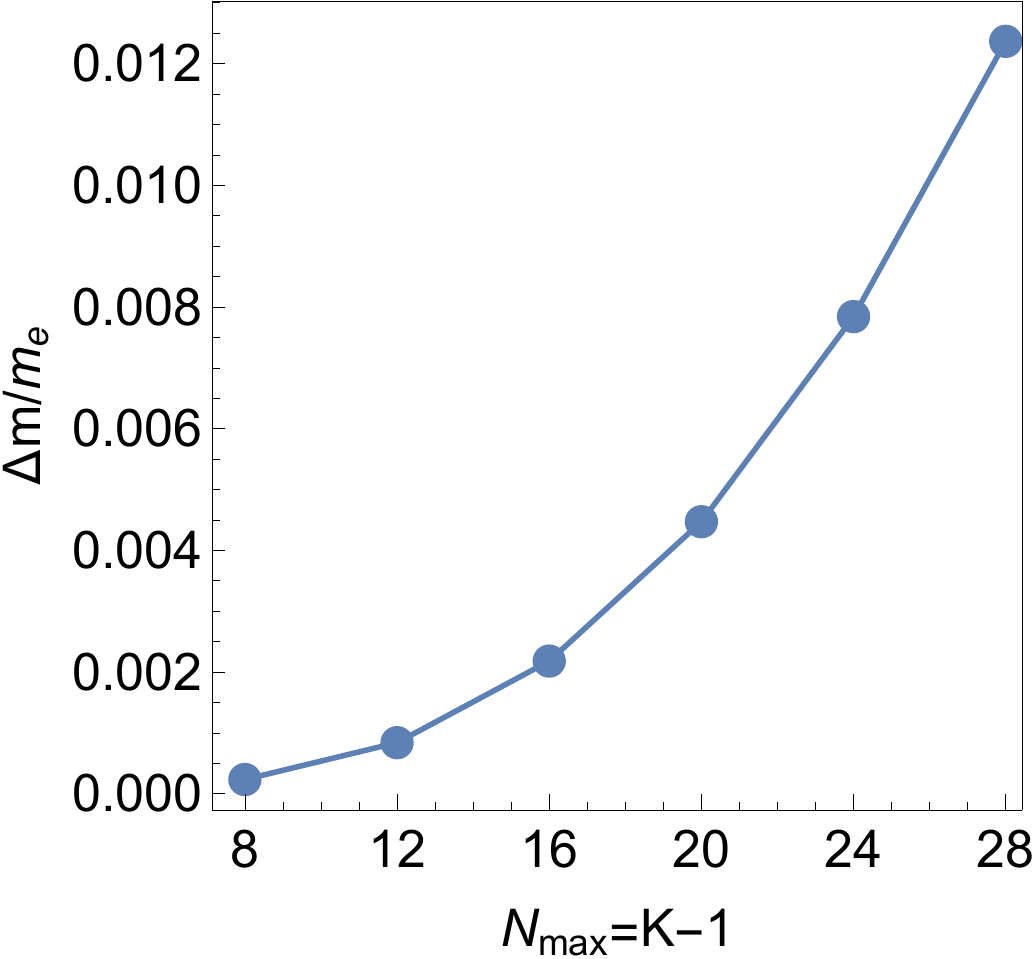}
\includegraphics[width=0.45\columnwidth
]{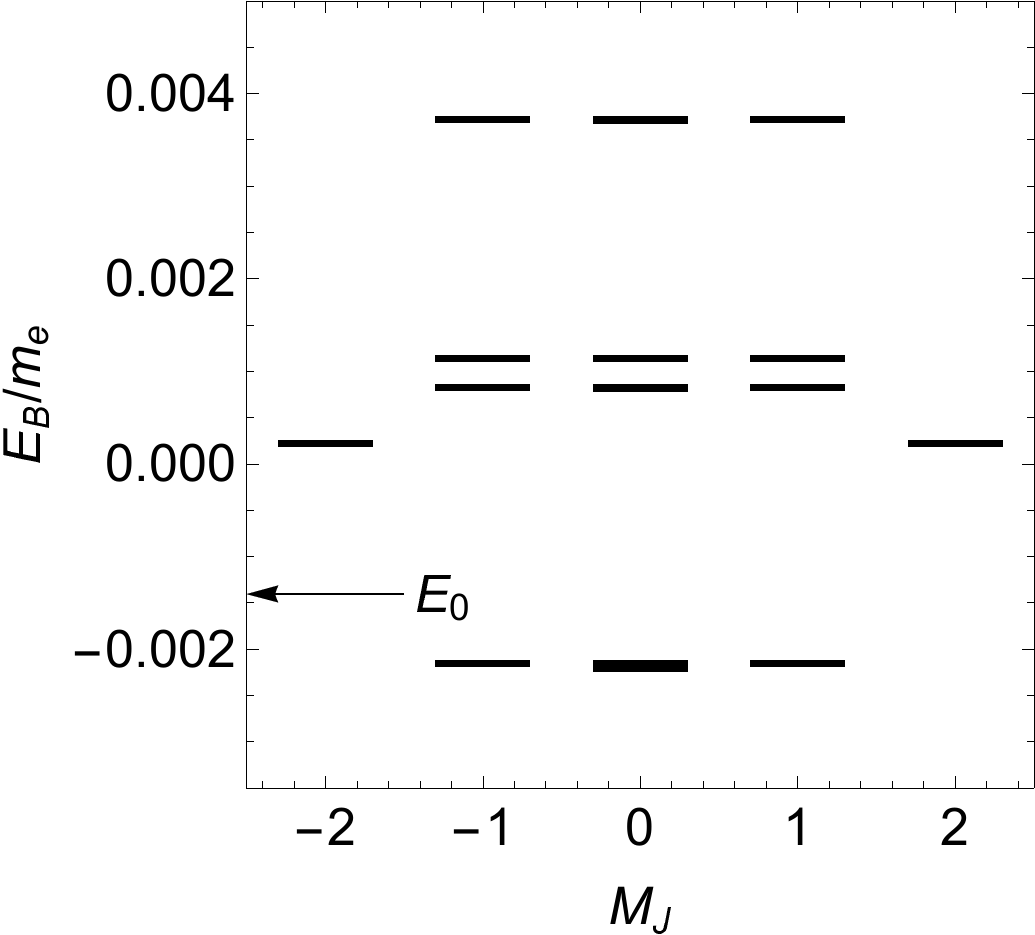}
\caption{\label{fig:massct} Left panel: representative value of the mass counterterm (in units of the physical electron mass $m_e$) in the positronium problem as a function of basis truncation parameters $N_{\rm{max}}=K-1$; right panel: the binding energy ($E_B$) spectrum of the positronium system at $N_{\rm{max}}=K-1=28$ and $\alpha=0.075$. $E_0$ is the ground state ($1^1S_0$) binding energy from nonrelativistic quantum mechanics with perturbative corrections. }
\end{figure}

In this calculation, we adopt the Fock-sector dependent renormalization~\cite{Zhao:2014hpa,Karmanov:2008,Karmanov:2012}, according to which only the fermion mass in the $|e^+ e^-\rangle$ sector needs to be renormalized, namely, the bare mass $m_{e0}$ is different from the physical mass $m_e$.  Different basis states take distinct values for the mass counterterm depending on their respective quanta available for self-energy fluctuation. The mass counterterms are determined from solving a series of single electron systems in the $|e\rangle+ |e \gamma\rangle$ Fock sectors~\cite{Zhao:2014xaa}. 
% The mass counterterm is a function of quantum numbers of the electron or positron in $|e^+ e^-\rangle$, where the positron has the same mass counterterm with the electron of the same quantum number. 

\begin{figure}
\centering
\includegraphics[width=0.3\columnwidth
]{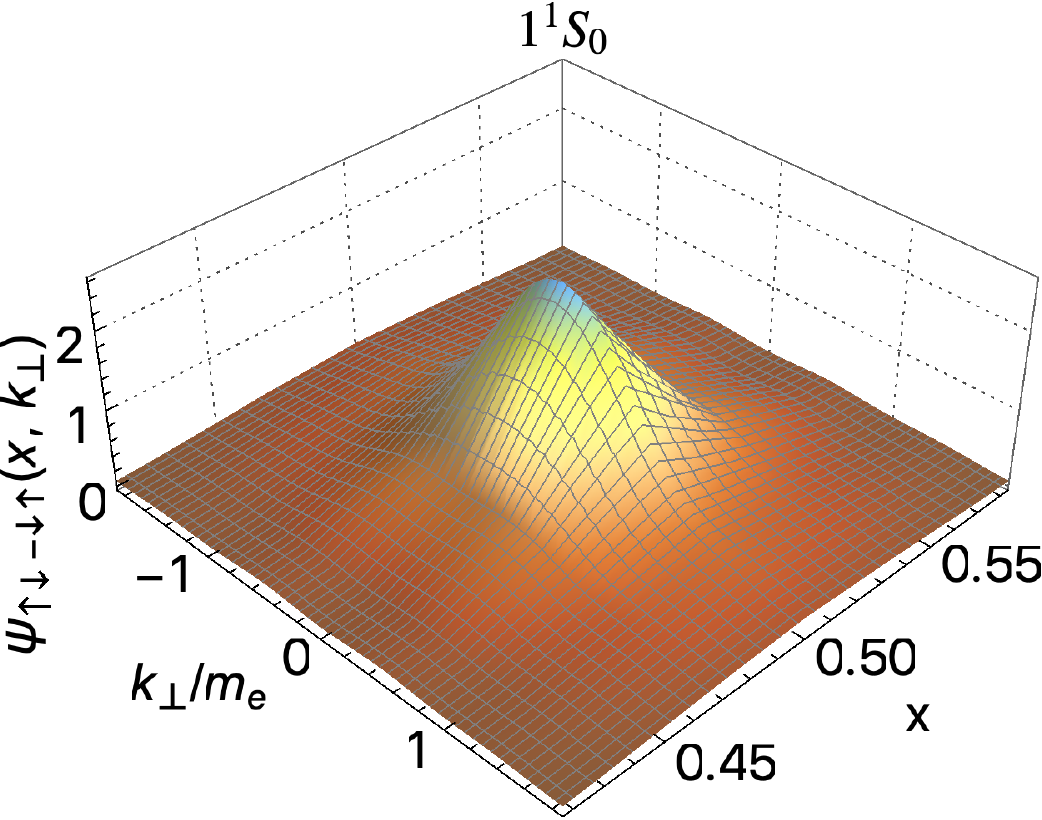}
\includegraphics[width=0.3\columnwidth
]{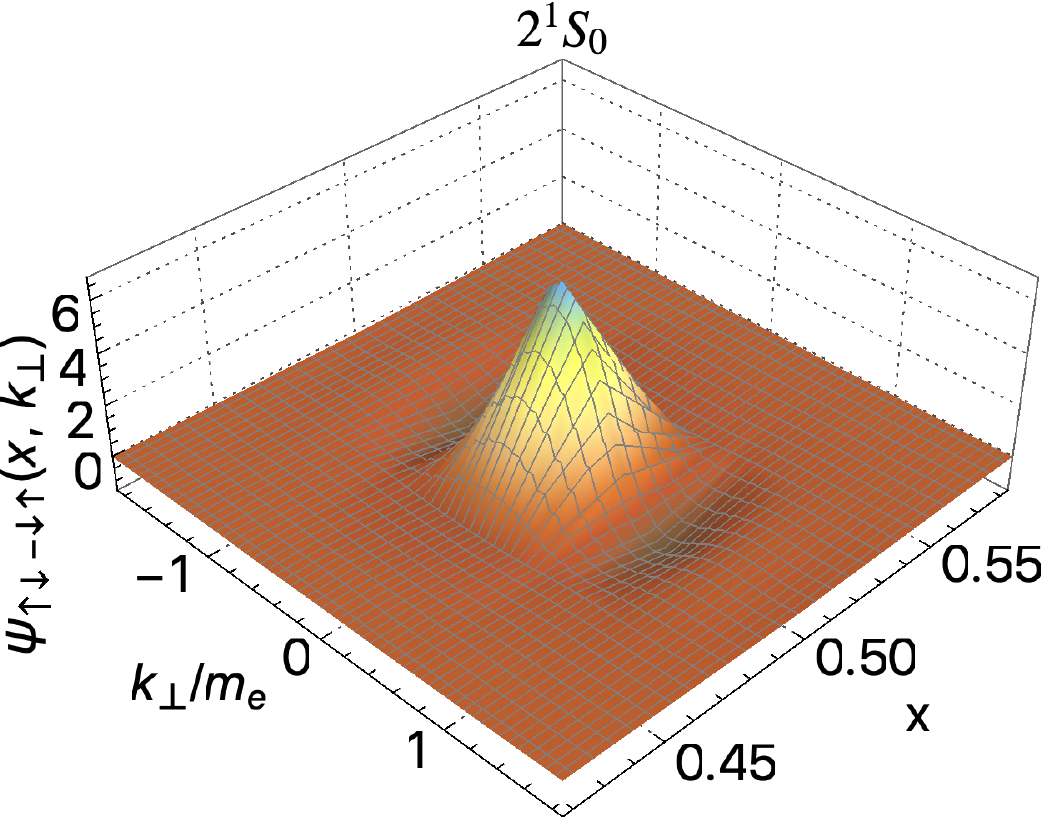}
\includegraphics[width=0.3\columnwidth
]{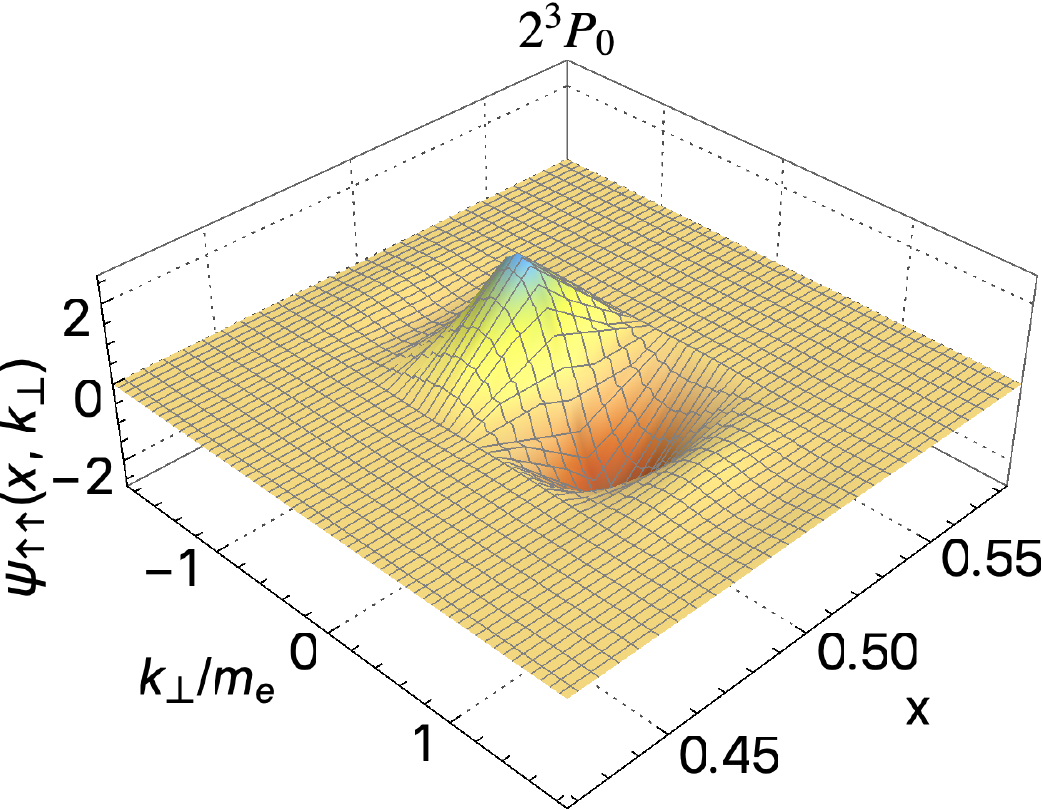}
\caption{\label{fig:wf} The (normalized) LFWFs for the dominant spin component in the $|e^+ e^-\rangle$ sector of the positronium system at $N_{\rm{max}}=K-1=28$ and $\alpha=0.075$. $x$ is the longitudinal momentum fraction and $k_\perp$ represents the relative transverse momentum between $e^+$ and $e^-$.}
\end{figure}

The value of the mass counterterm $\Delta m=m_{e0}-m_e$ for a representative basis state in the positronium problem as a function of the truncation parameters is shown in the left panel of Fig.~\ref{fig:massct} and in the right panel, we present the binding energy spectrum of the positronium system, $E_B \equiv M_{\rm{Ps}}-2\,m_e$, for different spin projections $M_J$. The arrow indicates the value of the ground state binding energy from nonrelativistic quantum mechanics~\cite{bs}, which is close to our result. We note that the mass counterterm is typically on a larger scale than that of the binding energy. We also note that the scale of the binding energy and structures of multiplets in the spectrum are in reasonable agreement with the previous calculation based on an effective one-photon-exchange interaction between the $e^+$ and $e^-$~\cite{Wiecki:2014}. The approximate degeneracy among different $M_J$ substates and the information from the mirror parity and charge parity~\cite{Li:2017mlw} allow us to identify the low-lying eigenstates. For the $M_J=0$ states, from bottom to up, the lowest six states are $1^1S_0$, $1^3S_1$, $2^1S_0$, $2^3S_1$, $2^3P_0$, and $2^3P_1$. Fig.~\ref{fig:wf} illustrates the light-front wave function (LFWF) in the $|e^+ e^-\rangle$ sector for three low-lying states with $M_J =0$. Their shape and nodal structures are qualitatively similar to those based on the one-photon-exchange effective interaction~\cite{Wiecki:2014}.

\section{Heavy meson}
A similar calculation can be carried over to the heavy meson system in QCD. Like the positronium system, we retain the lowest two Fock sectors, $|q\bar{q}\rangle$ and $|q\bar{q}g \rangle$, in our basis. Our Hamiltonian contains two parts. From the QCD Lagrangian, we obtain the first part of the Hamiltonian~\cite{Brodsky:1997de}, $P^-_{\rm{QCD}}$, which in our truncated Fock space takes a similar form to the QED Hamiltonian, $P^-_{\rm{QED}}$, in Eq.~(\ref{QEDHami}), with the fermion field $\Psi$ identified as the quark field, the gauge boson field $A_\mu$ identified as the gluon field $A^a_\mu T^a$ and the electric charge $e$ replaced by the color charge $g$. In order to achieve a more accurate reproduction of the meson mass spectrum, we allow the quark mass appearing in the quark-gluon vertex interaction, $gj^\mu A^a_\mu T^a$, to be an independent phenomenological parameter, $m'_q$, from $m_q$ in the kinetic energy term. We also apply a nonzero gluon mass $m_g$ to ensure the low-lying states are dominated by the $|q\bar{q}\rangle$ sector.
% \begin{equation}
% \begin{split}
% P_{QCD}=&\int d^2x^{\perp}dx^{-} \frac{1}{2}\bar{\Psi}\gamma^+ \frac{m^2+(i\partial^{\perp})^2}{i\partial^+} \Psi -\frac{1}{2}A^{i}_{a}(i\partial^{\perp})^2A^{i}_{a}\\
% &+g\bar{\Psi}\gamma_{\mu}A^{\mu}\Psi+\frac{1}{2}g^2 \bar{\Psi}\gamma^{+}T^{a}\Psi\frac{1}{(i\partial)^2}\bar{\Psi}\gamma^{+}\Psi,
% \end{split}
% \end{equation}
% where the first line of $P_{QCD}$ is the kinetic term, the second line include the vector coupling vertex and the vector coupling with instantaneous gluon. $m$ is the mass of the quark (anti-quark). $\Psi$ and $A_{\mu}$ are the fermion and gluon fields\\
 
In addition we include a phenomenological confining potential~\cite{Li:2017mlw} in both the longitudinal and transverse directions in the $|q\bar{q}\rangle$ sector, which takes the following form, 
\begin{equation}
P_{\rm C}^-P^+=\kappa^{4} \vec{\zeta}_{\perp} - \frac{\kappa^4}{(m_{q}+m_{\bar{q}})^2}\partial_{x}(x (1-x)\partial_{x}),
\end{equation}
where $x$ is the longitudinal momentum fraction of the quark~\cite{Wiecki:2014}. $\vec{\zeta}_{\perp}\equiv \sqrt{x(1-x)} \vec{r}_\perp$ is the holographic variable introduced by Brodsky and de T\'eramond~\cite{Brodsky:2014yha}, and $\partial_x f(x, \vec\zeta_\perp) = \partial f(x, \vec \zeta_\perp)/\partial x|_{\vec\zeta}$. $\kappa$ is the strength of the confinement and $m_{q}(m_{\bar{q}})$ is the mass of the quark (anti-quark). Thus, our total Hamiltonian is $P^-=P^-_{\rm QCD}+P^-_{\rm C}$.
% We only add the confining potential to the leading Fock sector $|q\bar{q}\rangle$, because it is complicated to the interaction between the quark and the gluon. 

%  So, we get the light-front Hamiltonian formalism $P^-=P^-_{QCD}+P^-_{V}$. In principle, hadron observables can be evaluated by solving the eigenvalue equation\cite{Wiecki:2014}
%  \begin{equation}
%  P^{\mu}P_{\mu}|\Psi\rangle = M^2|\Psi \rangle,
%  \end{equation}
% where $P^{\mu}$ is the energy-momentum four-vector operator. $M$ is the invariant mass of the quarkonium. we adopt the Fock sector dependent renormalization~\cite{Zhao:2014hpa,Karmanov:2008,Karmanov:2012}to the quarkonium calculation like the positronium. 

\begin{figure}
\centering
\includegraphics[width=0.32\columnwidth
]{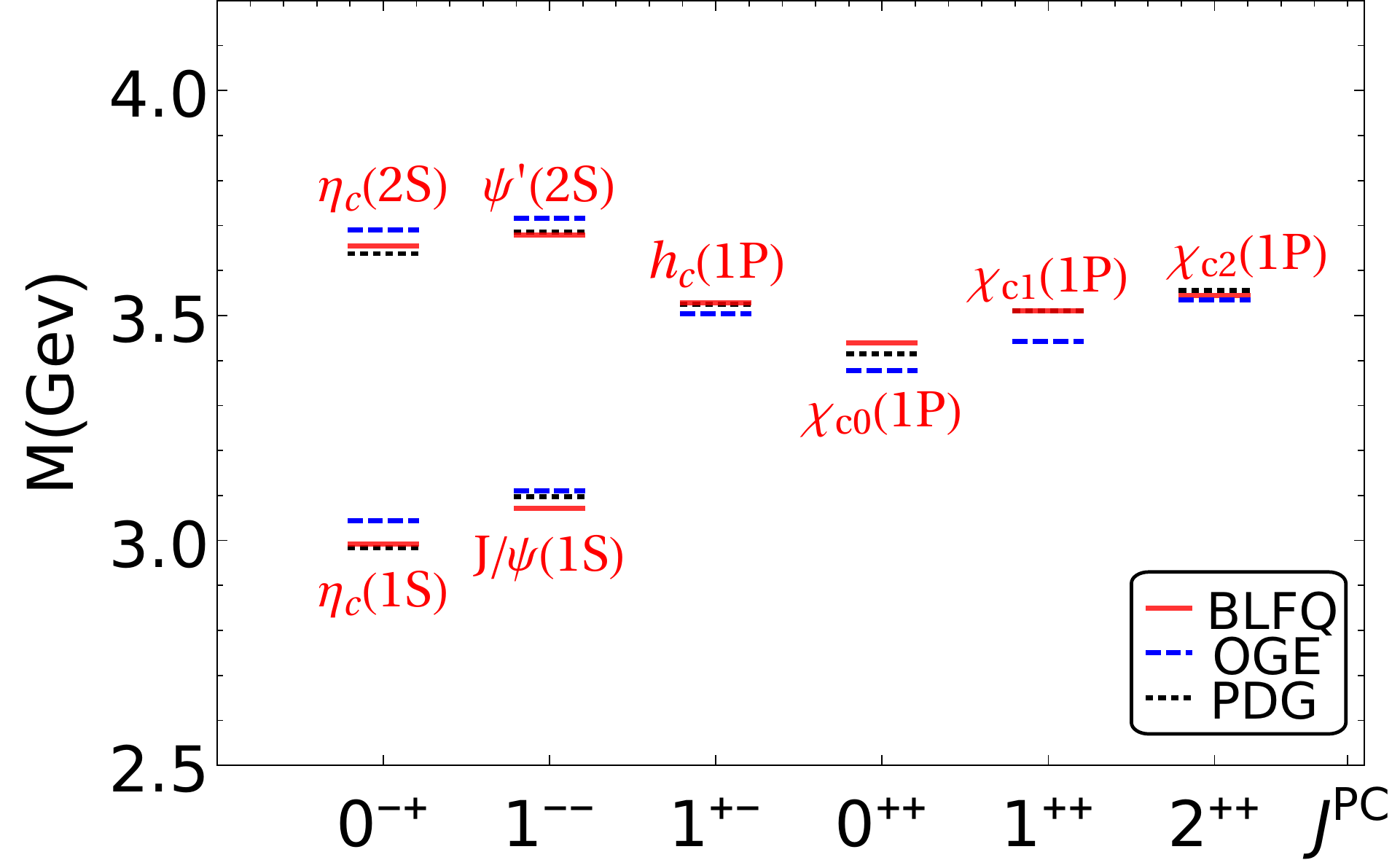}
\includegraphics[width=0.32\columnwidth
]{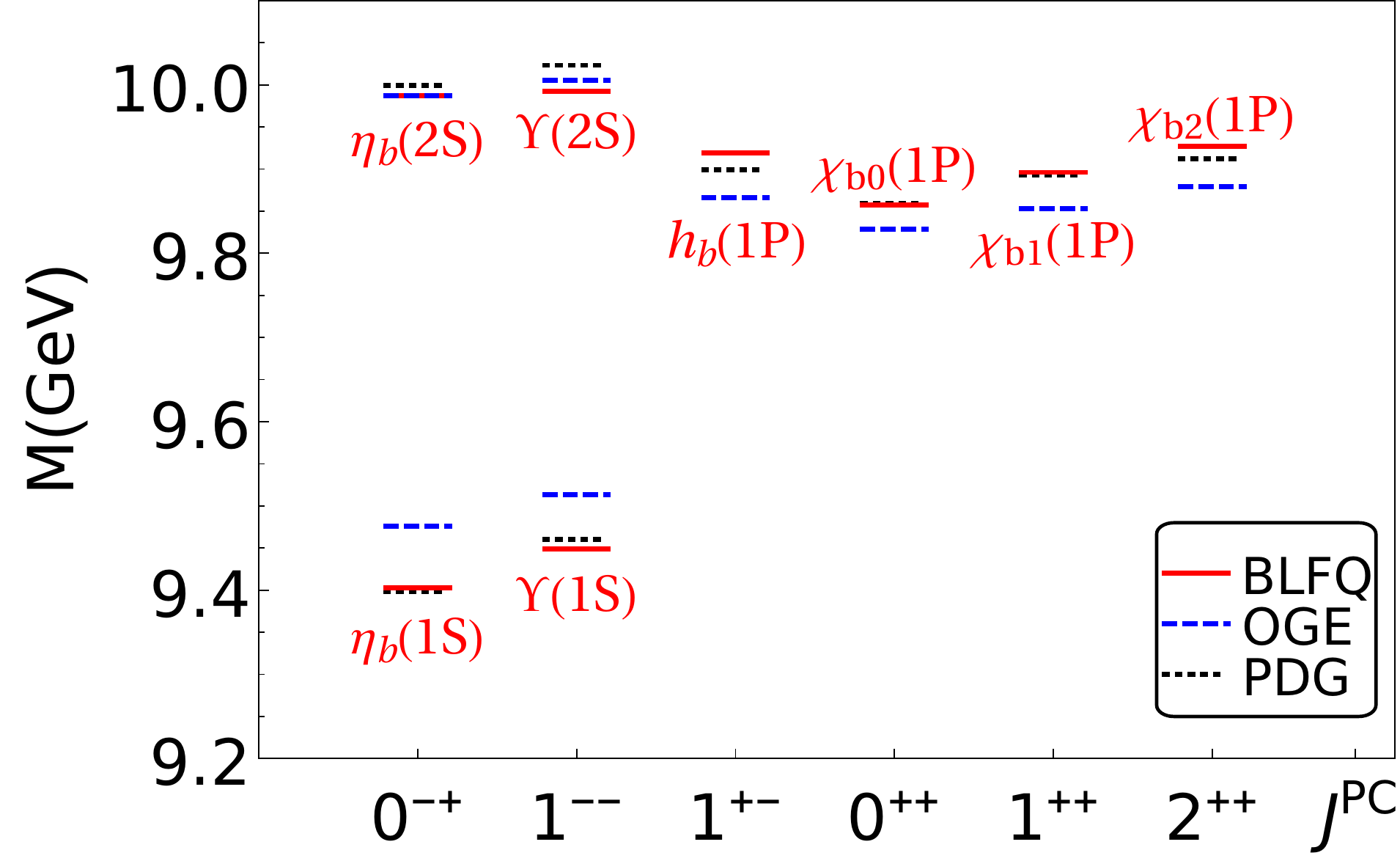}
\includegraphics[width=0.32\columnwidth
]{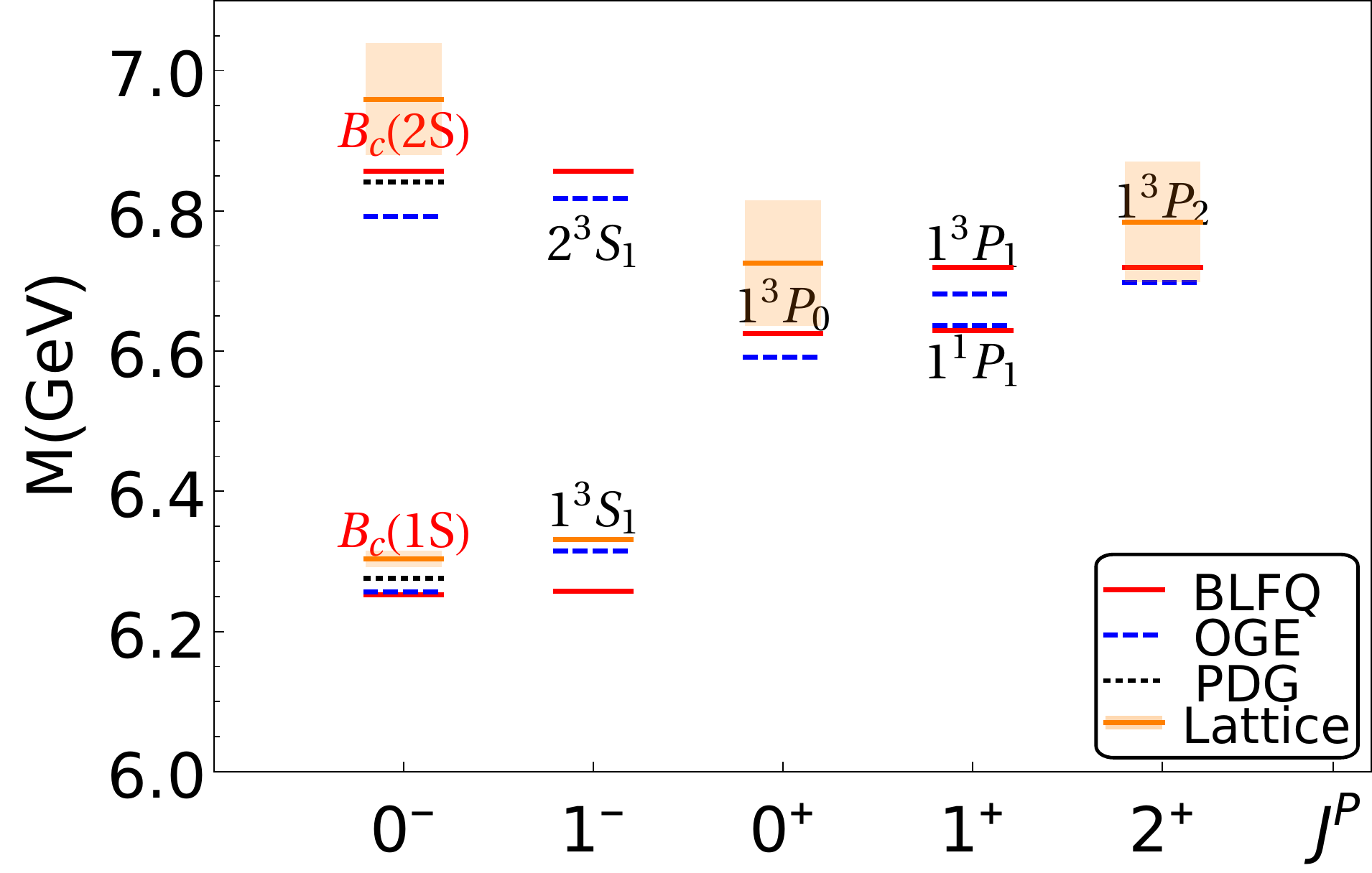}
\caption{\label{fig:en} Comparison of our BLFQ spectra at $N_{\rm{max}}=K-1=8$ for charmonium (left), bottomonium (middle), and $B_c$ meson (right) with the effective one-gluon-exchange (OGE) approach~\cite{Li:2017mlw,Tang:2018myz} and the experimental values (PDG)~\cite{Tanabashi:2018oca}. Lattice results are from Ref.~\cite{Allison:2004be,Gregory:2009hq,Davies:1996gi}. The horizontal and vertical axes are the $J^{\mathsf P\mathsf C}$ and invariant mass, respectively. }
\end{figure}

\begin{figure}
\centering
\includegraphics[width=0.3\columnwidth
]{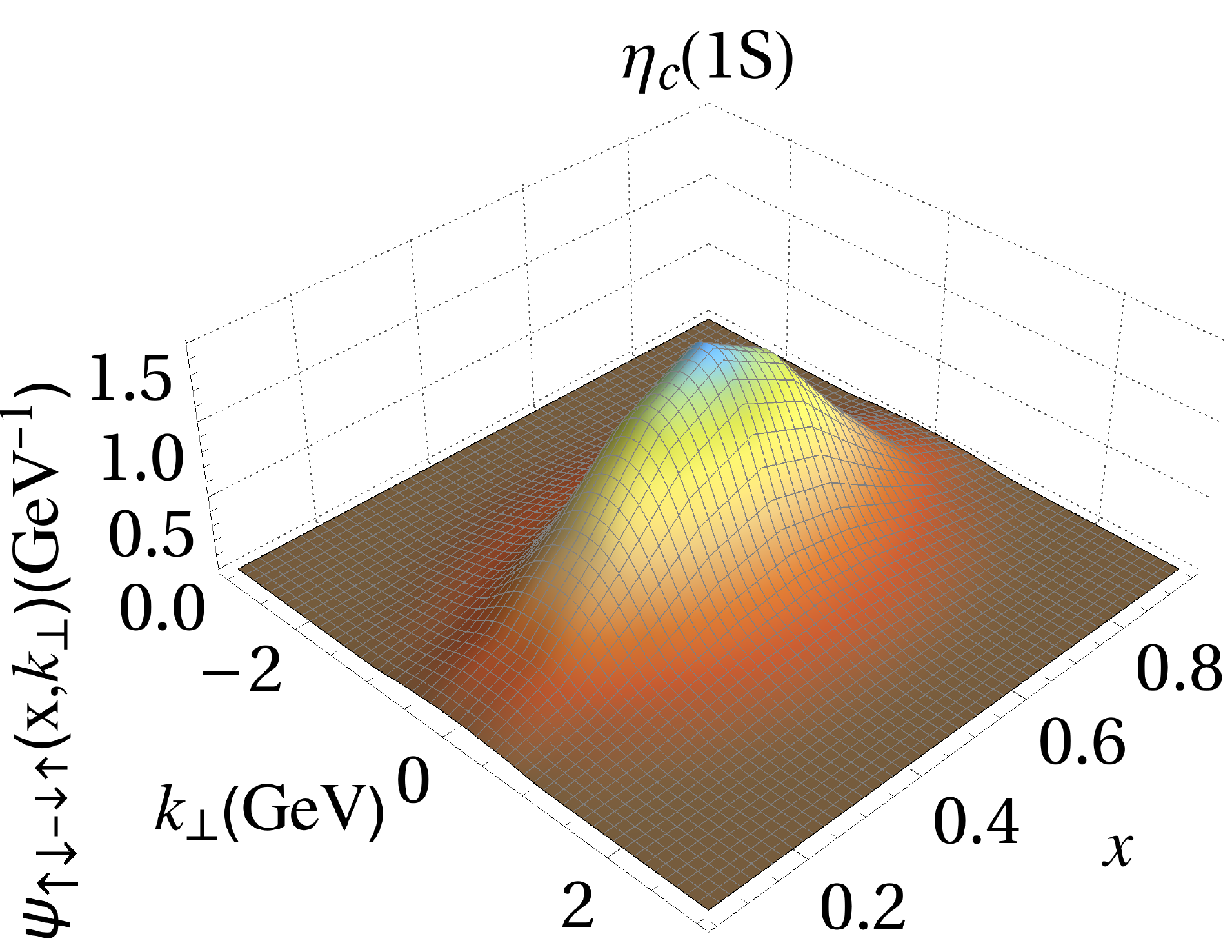}
\includegraphics[width=0.3\columnwidth
]{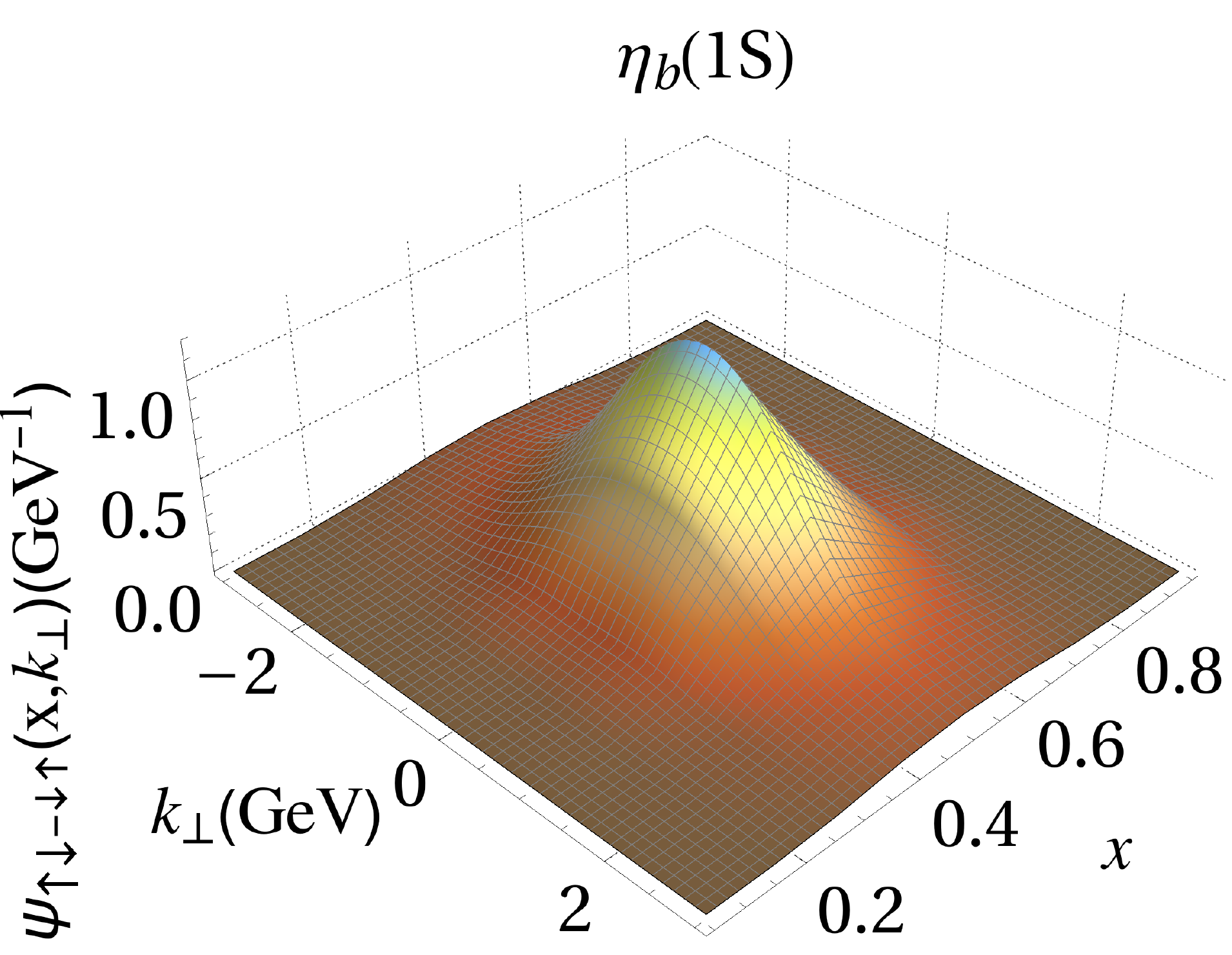}
\includegraphics[width=0.3\columnwidth
]{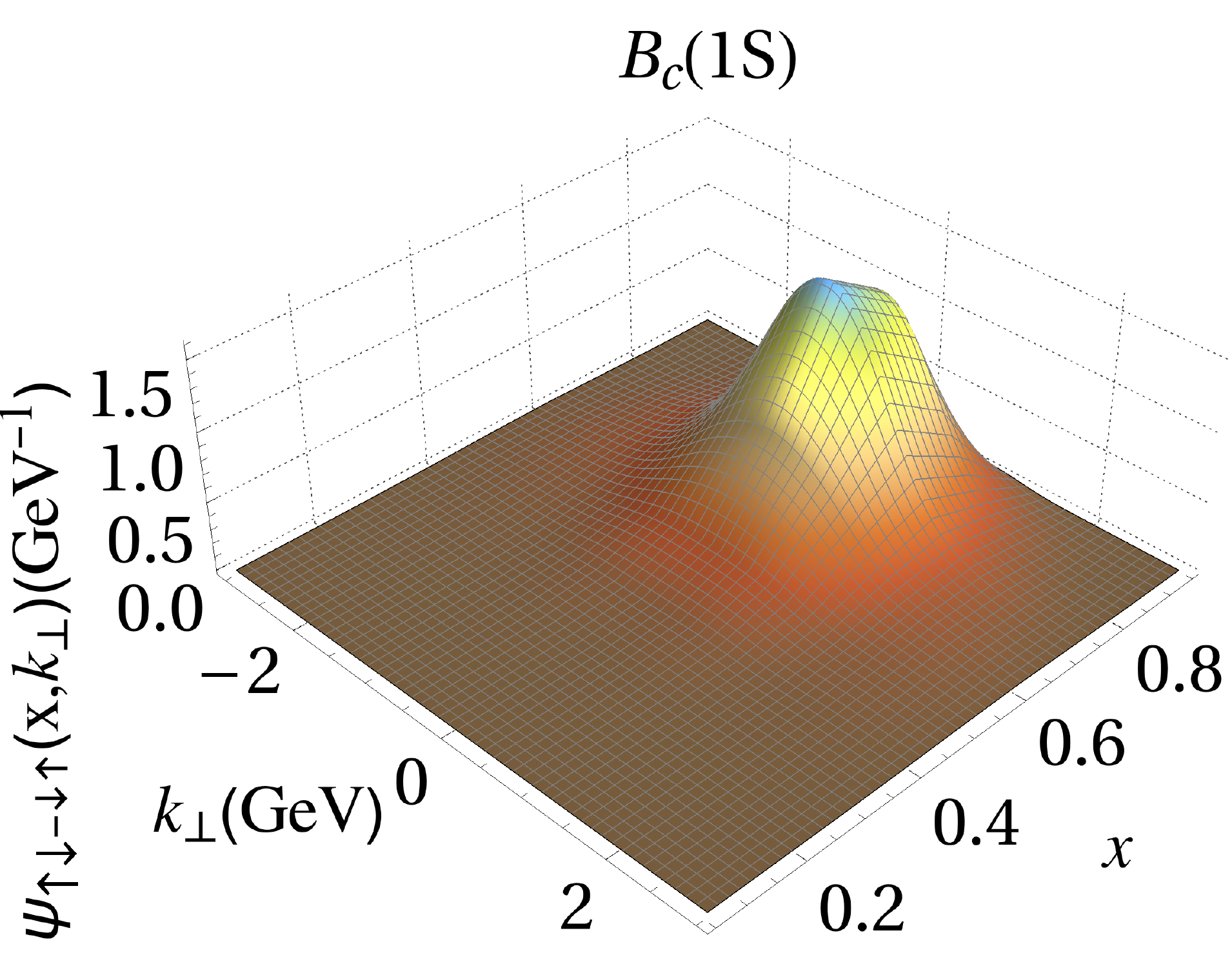}
\caption{\label{fig:wfq1} The (normalized) LFWFs for the dominant spin component in the $|q\bar{q}\rangle$ sector of $\eta_c(1S)$ (left), $\eta_{b}(1S)$ (middle) and $B_c(1S)$ (right) at $N_{\rm{max}}=K-1=8$. $x$ is the longitudinal momentum fraction of the quark and $k_\perp$ represents the relative transverse momentum between the quark and the anti-quark.}
\end{figure}

% We get the invariant mass of the charmonium and bottomonium by fitting experiments (PDG). 
With the charm quark mass $m_c=1.561$\,GeV (kinetic), $m'_c=4.343$\,GeV (interaction), the gluon mass $m_g=0.4$\,GeV, the confining strength $\kappa=1.14$\,GeV and the strong coupling constant $g=2.3$ for the charmonium, and similarly the bottom quark mass $m_b=4.767$\,GeV (kinetic), $m_f=15$\,GeV (interaction), $m_g=0.4$\,GeV, $\kappa=1.948$\,GeV and $g=1.6$ for the bottomonium, our resulting mass spectra for the low-lying $c\bar{c}$ and $b\bar{b}$ states agree with the experimental values reasonably well, as shown in the left and middle panel of Fig.~\ref{fig:en}, respectively. With the same parameter set used in the $c\bar{c}$ and $b\bar{b}$ systems we obtain the mass spectrum for the $B_c$ system, as in the right panel of Fig.~\ref{fig:en}. % We only renormalize the $|q\bar{q}\rangle$ Fock sector, then plot the wave function of the quarkonium in the leading Fock sector.
In Fig.~\ref{fig:wfq1} we present the ground state LFWFs of the $c\bar{c}$, $b\bar{b}$ and $B_c$ system in the $|q\bar{q}\rangle$ sector.

% \begin{figure}
% \centering
% \includegraphics[width=0.3\columnwidth
% ]{figures/ccbar_state2}
% \includegraphics[width=0.3\columnwidth
% ]{figures/bbbar_state2}
% \includegraphics[width=0.3\columnwidth
% ]{figures/bcbar_state2}
% \caption{\label{fig:wfq2} LFWFs of $J/ \psi(1S) \quad \Upsilon_{b}(1S)$ and $B_c(1^{3}S_{1})$ at $N_{\rm{max}}=K-1=8$ in the $|q^+ q^-\rangle$ Fock sector.}
% \end{figure}

\section{Light meson}\label{aba:sec2}
\begin{figure}
\begin{center}
\includegraphics[width=2.25in]{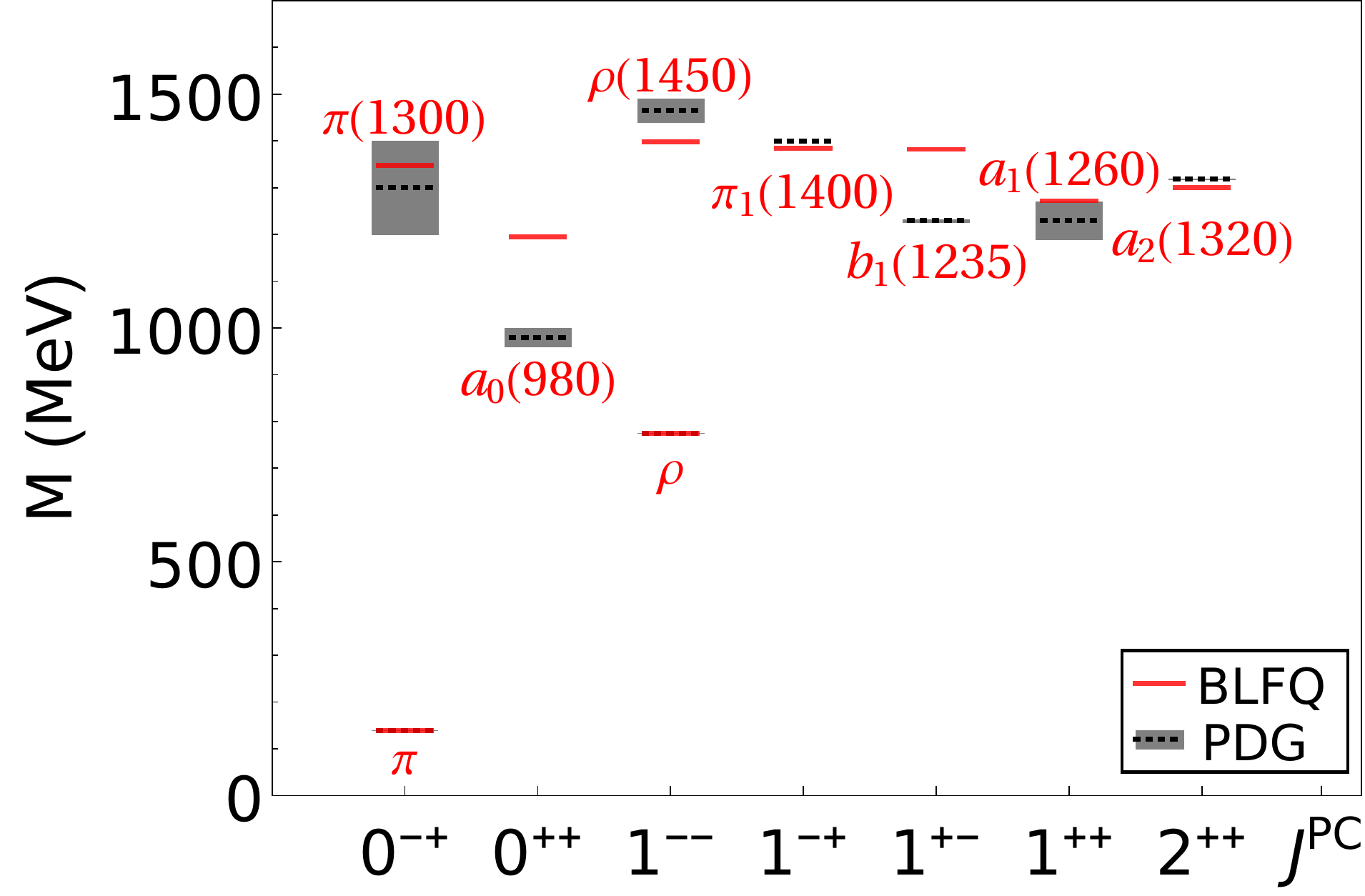}
\includegraphics[width=2.2in]{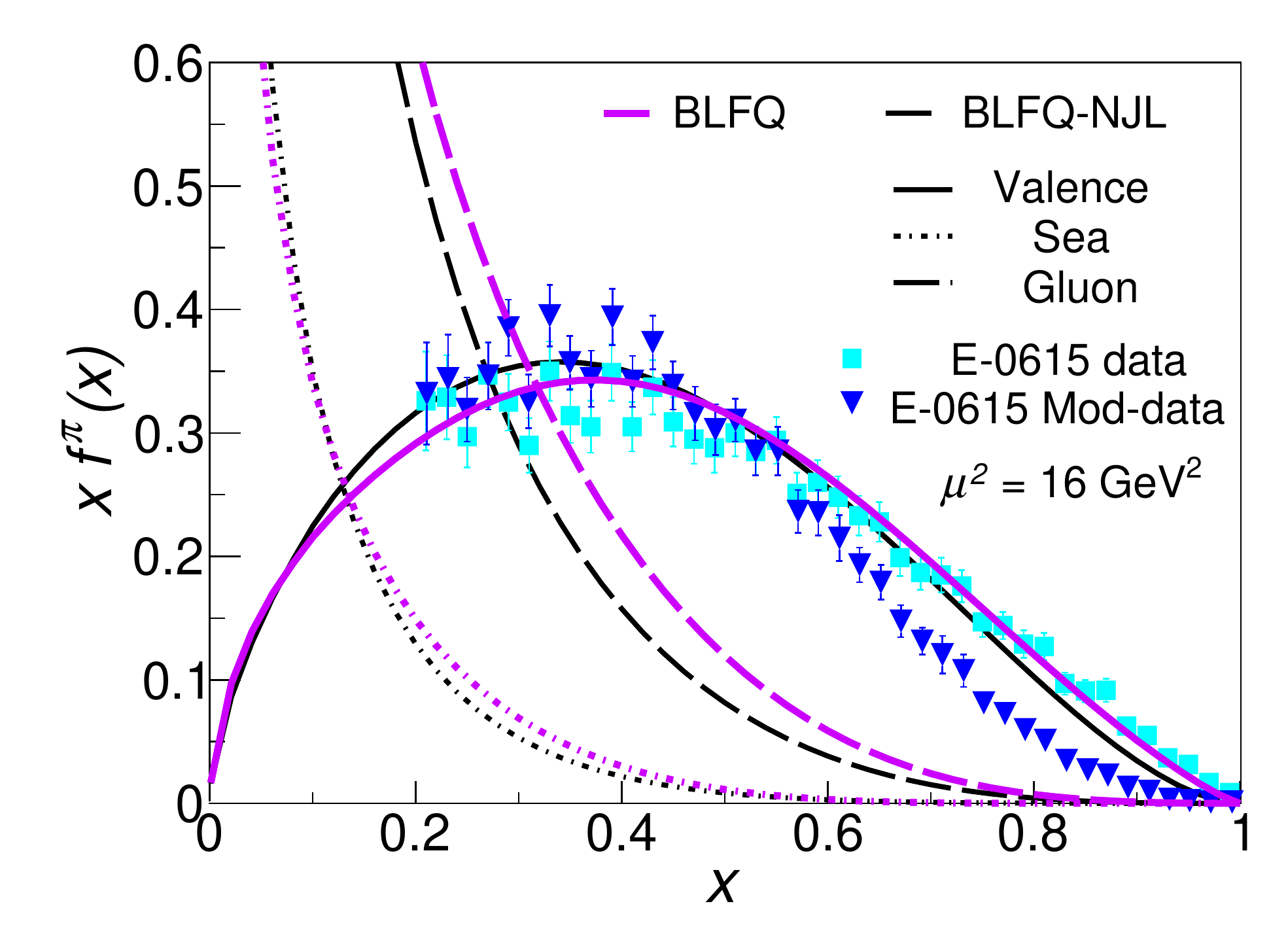}
\caption{Left panel: the mass spectrum of the light mesons; right panel: longitudinal momentum distribution $xf^\pi(x)$ as a function of longitudinal momentum fraction $x$ of a parton in the pion.}
\label{fig1}
	\end{center}
\end{figure}

We perform a similar calculation in the light meson sector. With $m_u=m_d=0.33$\,GeV, $m'_u=m'_d=3.38$\,GeV, $\kappa=0.77$\,GeV and $g$=1.74, we obtain the mass spectrum of the light mesons, as shown in the left panel of Fig.~\ref{fig1}.
% In light mesons calculation, we truncated Fock space containing up to next-to-order Fock sector, $|\pi^+\rangle= a|u\bar{d}\rangle+ b|u\bar{d}g\rangle$.
% We adopt Hamiltonian, including the light-front QCD Hamiltonian as well as a transverse\cite{Brodsky:2014yha} and a longitudinal\cite{Li:2017mlw} confining potentials, $H =P^{-}_{QCD}+P_V$, to solve the time-independent light-front Schr\"{o}dinger equation, $H\vert \Psi\rangle=M^2\vert \Psi\rangle$. Here, the light-front QCD Hamiltonian are given by \cite{Brodsky:2014yha},
% \begin{equation}
% \begin{split}
% P^-_{QCD}=&\int d^2 x^{\perp} dx^- \frac{1}{2} \bar{\Psi} \gamma^+ \frac{m^2+(i\partial^{\perp})^2}{i\partial^+} \Psi -\frac{1}{2}A^i_a (i\partial^{\perp})^2 A^i_a \\
% &+g \bar{\Psi} \gamma_{\mu} A^{\mu} \Psi +\frac{1}{2}g^2 \bar{\Psi} \gamma^+ T^a \Psi \frac{1}{(i\partial^+)^2} \bar{\Psi} \gamma^+ T^a \Psi,
% \end{split}
% \end{equation}
% where, the first two items are the light-front kinetic energy for the quark, the antiquark, and the gluon, while  the last two items are the vertex interaction and the instantaneous gluon
% exchange interaction, respectively.

% In Figure~\ref{fig1} (left panel), we show the mass spectrum of the light mesons. 
Although the masses for $a_0(980)$ and $b_1(1235)$ somewhat deviate from the experimental value~\cite{Tanabashi:2018oca}, the results for $\pi$, $\rho$, $a_1(1260)$, $\pi(1300)$, $a_2(1320)$, and $\pi_1(1400)$ are in good agreement with the experimental data~\cite{Tanabashi:2018oca}. 

Next, we calculate the pion's parton distribution function (PDF), which represents the probability density of finding a parton with the longitudinal momentum fraction $x$ inside the pion. We first calculate the initial scale PDF based on the LFWF from BLFQ in both the $|q\bar{q}\rangle$ and $|q\bar{q}g\rangle$ sectors.
% \begin{equation}
% f^\pi(x)=\int \frac{d^2\vec{p}_{\perp}}{(2\pi)^2} \psi^*_\pi(\vec{p}_{\perp},x) \psi_\pi(\vec{p}_{\perp},x),
% \end{equation}
% where $\psi_\pi$ is the normalized LFWF in the $|q^+ q^-\rangle$ sector. 
We then evolve the initial scale PDF from ${\mu_0^2=0.31~\mathrm{GeV}^2}$ to the relevant experimental scales ${\mu^2=16~\mathrm{GeV}^2}$ by the DGLAP equations~\cite{Dokshitzer:1977sg,Gribov:1972ri,Altarelli:1977zs} using the higher order perturbative parton evolution toolkit~\cite{Salam:2008qg}. 

In Fig.~\ref{fig1} (right panel), we compare our results with the E615 experimental result~\cite{Conway:1989fs} and the modified result of the E615 experiment~\cite{Chen:2016sno}. We find that, for the valence PDF, our result shows slightly better agreement with the E615 experimental result~\cite{Conway:1989fs} in the high-$x$ region compared to the earlier calculation based an effective Nambu–Jona-Lasinio interaction (BLFQ-NJL). For the gluon PDF, our result is larger than that obtained from the BLFQ-NJL model, and for the sea PDF, our result is close to that obtained from the BLFQ-NJL model~\cite{Lan:2019vui}.

\section{Baryon}
\begin{figure*}[htbp]
\centering
\includegraphics[width=2.2in]{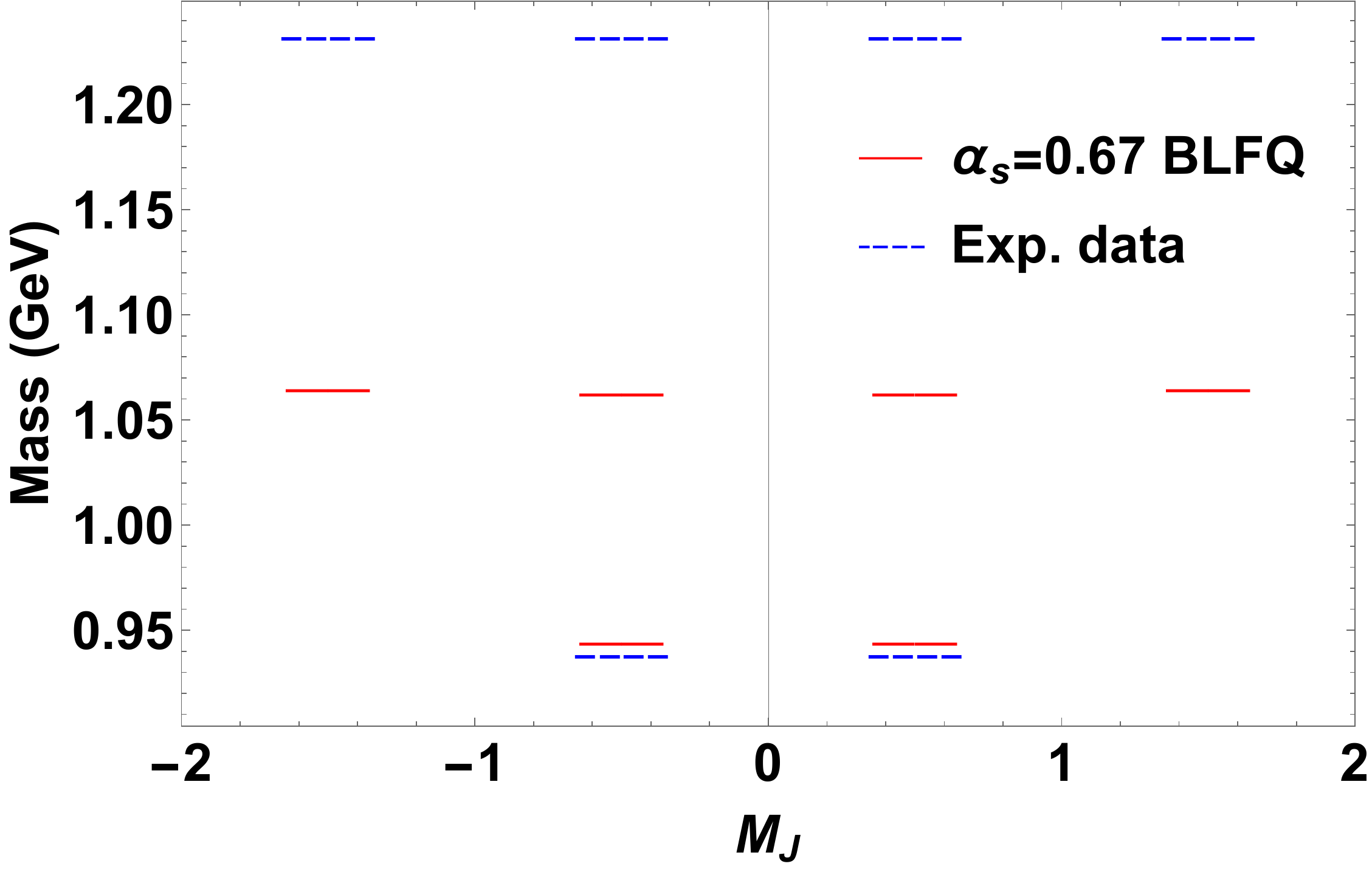}
\includegraphics[width=2.2in]{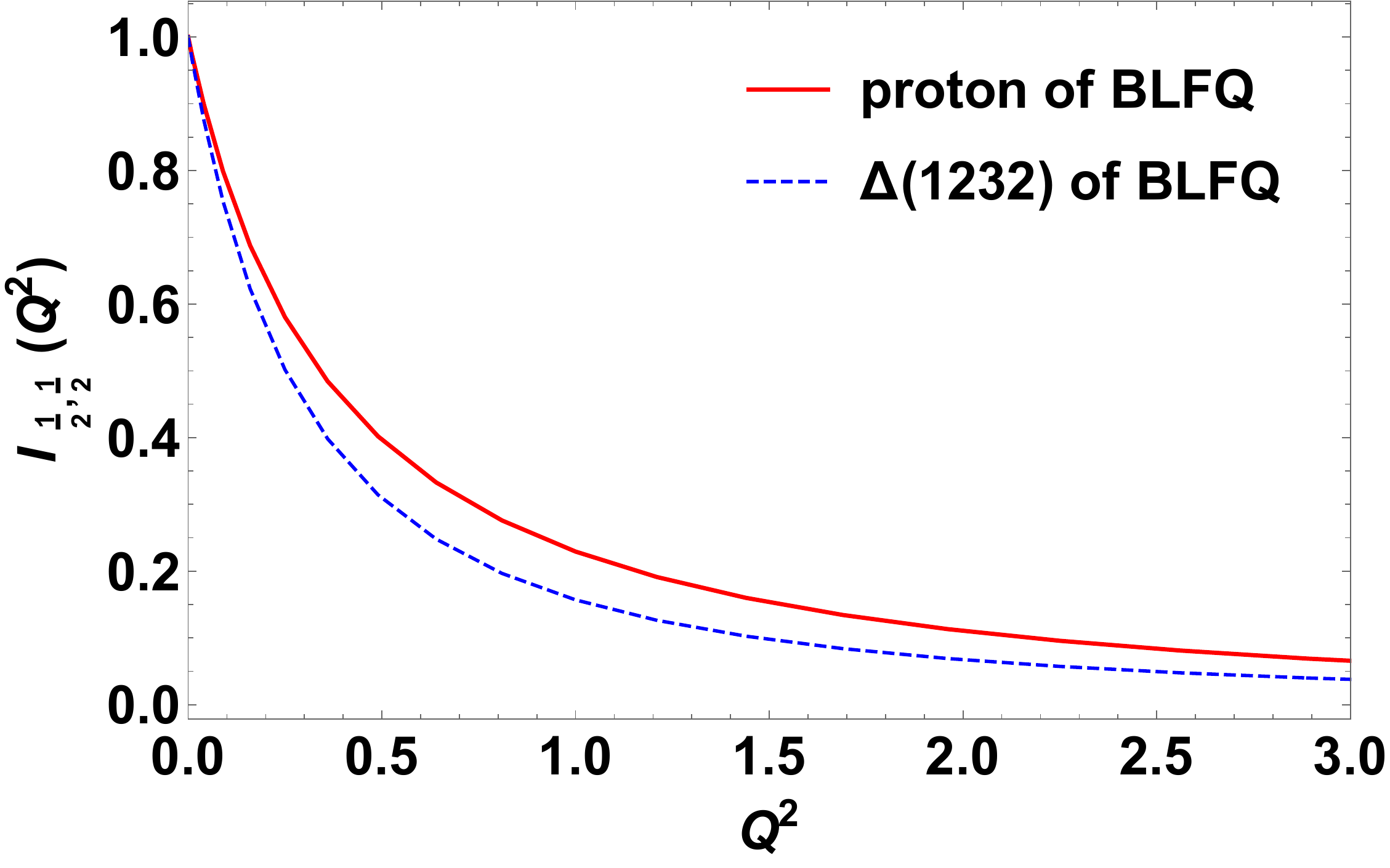}
\caption{Left panel: the masses of the nucleon and $\Delta(1232)$ compared to experimental data~\cite{Tanabashi:2018oca}; right panel: comparison of the $I_{\frac{1}{2},\frac{1}{2}}(Q^2)$ elastic form factor between the proton and $\Delta^+$. }
\label{delta}
\end{figure*}

We perform an initial calculation of the baryon system based on an effective interaction in the $|qqq\rangle$ sector~\cite{Xu:2019xhk,Xu:2020}. This effective interaction consists of the (pairwise) one-gluon-exchange interaction and the (pairwise) longitudinal and transverse confining interactions. The parameters in the effective interaction are fixed by the mass and Dirac form factor of $u, d$ flavor. For the baryon system, the ground state spin-1/2 (3/2) particle is the nucleon ($\Delta~(1232)$). Our preliminary results of their masses are compared with the experimental data in the left panel of Fig.~\ref{delta}. Our mass of $\Delta~(1232)$ is about 170\,MeV smaller than the experimental value. It remains to be seen that whether this mass discrepancy will decrease as the basis size increases. We also calculate the $I_{\frac{1}{2},\frac{1}{2}}(Q^2)$ elastic form factor~\cite{Adhikari:2016idg} of the proton and $\Delta^+$, which is compared in Fig~\ref{delta}. The larger slope of $I_{\frac{1}{2},\frac{1}{2}}$ for $\Delta^+$ at $Q^2\to 0$ suggests that the charge radius of $\Delta^+$ is larger than that of the proton, which is as expected since $\Delta^+$ is an excitation of the proton.
% For $\Delta^+$ we actually calculate the form factor corresponding to MOur calculation of the system mass is less than the experimental data. We compare the Dirac form factors of the $\Delta$ with the proton system, where it shows the charge radius of the $\Delta$ is less than the proton~\cite{Tanabashi:2018oca}. This conclusion agrees with the lattice results~\cite{Alexandrou:2007we}.

\section{Conclusion}
Through the applications to various bound state systems in QED and QCD we demonstrate that BLFQ is a versatile and powerful nonperturbative approach to quantum field theory for strongly interacting systems. We anticipate that in the future BLFQ will be a useful tool for understanding hadron mass spectrum and structure beyond the valence sector.

\section{Acknowledgment}
XZ is supported by Key Research Program of Frontier Sciences, CAS, Grant No ZDBS-LY-7020. CM is supported by the National Natural Science Foundation of China (NSFC) under the Grant No. 11850410436 and No. 11950410753. JPV is supported by the Department of Energy under Grants No. DE-FG02-87ER40371, and No. DE-SC0018223 (SciDAC4/NUCLEI). A portion of the computational resources were provided by the National Energy Research Scientific Computing Center (NERSC), which is supported by the Office of Science of the U.S. Department of Energy under Contract No.DE-AC02-05CH11231.

\end{document}